\documentclass[iop]{emulateapj}

\usepackage{amsmath,amssymb,times,graphics}

\newcommand{\masyr}{mas~yr\ensuremath{^{-1}}}

\newcommand{\Msun}{M\ensuremath{_{\sun}}}
\newcommand{\slowpokes}{SLoWPoKES}
\newcommand{\Pf}{P\ensuremath{_{\rm f}}}

\defcitealias{Dhital2010}{Paper I}

\shorttitle{{\slowpokes}--II}
\shortauthors{Dhital et al.}

\begin{document}

\journalinfo{{\it The Astronomical Journal}, accepted}
\submitted{Submitted: January 11, 2015; accepted: April 16, 2015; published:}

\title{{\slowpokes}-II: 100,000 Wide Binaries Identified in SDSS without Proper Motions}

\author{
  Saurav Dhital      \altaffilmark{1,2,3},
  Andrew A.\ West    \altaffilmark{2},
  Keivan G.\ Stassun \altaffilmark{3,4},
  Kyle J. Schluns    \altaffilmark{2},
  Angela P. Massey   \altaffilmark{2}}
\altaffiltext{1}{Department of Physical Sciences, Embry--Riddle Aeronautical University, 600 South Clyde Morris Blvd., Daytona Beach, FL 32114, USA; dhitals@erau.edu}
\altaffiltext{2}{Department of Astronomy, Boston University, 725 Commonwealth Avenue, Boston, MA 02215, USA.}
\altaffiltext{3}{Department of Physics \& Astronomy, Vanderbilt University, 6301 Stevenson Center, Nashville, TN, 37235, USA} 
\altaffiltext{4}{Department of Physics, Fisk University, 1000 17th Avenue N., Nashville, TN 37208, USA.}

\begin{abstract}
We present the {\slowpokes}--II catalog of low-mass visual binaries
identified from the Sloan Digital Sky Survey by matching photometric
distances. The candidate pairs are vetted by comparing the stellar
density at their respective Galactic positions to Monte Carlo
realizations of a simulated Milky Way. In this way, we are able to
identify large numbers of bona fide wide binaries without the need of
proper motions. 105,537 visual binaries with angular separations of
$\sim$1--20$\arcsec$, are identified, each with a probability of chance alignment of
$\leq$5\%. This is the largest catalog of bona fide wide binaries to date,
and it contains a diversity of systems---in mass, mass ratios, binary
separations, metallicity, and evolutionary states---that should
facilitate follow-up studies to characterize the properties of M
dwarfs and white dwarfs. There is a subtle but definitive suggestion
of multiple populations in the physical separation distribution,
supporting earlier findings. We suggest that wide binaries are
comprised of multiple populations, most likely representing different
formation modes. There are 141 M7 or later wide binary candidates,
representing a 7-fold increase in the number currently known. These
binaries are too wide to have been formed via the ejection
mechanism. Finally, we find that $\sim$6\% of spectroscopically confirmed M
dwarfs are not included in the SDSS \textsc{STAR} catalog; they are
misclassified as extended sources due to the presence of a nearby or
partially resolved companion. The {\slowpokes}--II catalog is publicly
available to the entire community on the world wide web via the
Filtergraph data visualization portal. 
\end{abstract}

\keywords{
(stars:) binaries: visual ---
stars: low mass ---
stars: brown dwarfs ---
stars: late-type ---
(stars:) white dwarfs ---
(stars:) subdwarfs
stars: statistics ---
}

\section{Introduction}\label{Sec:intro}
Components of binary (or multiple) systems are ideal coeval
laboratories to study star formation, to benchmark stellar
evolutionary models, and to calibrate empirical relations that determine
fundamental stellar parameters. While detailed and precise
measurements of individual objects or small samples provide important
tests for evolutionary models \citep[e.g.,][]{White1999,Stassun2007,Stassun2008}, 
large statistical samples are necessary for properly constraining the
behavior and intrinsic variation of star formation and of stellar
properties. All of these rest on the premise that individual
components in a stellar system are formed at
the same time \citep{White2001,Kraus2009a}, of the same material
\citep{Schuler2011,Dhital2012}, and have evolved in the same environment.

Until recently, the intrinsic faintness of low-mass stars---generally
defined as the regime bracketed by the hydrogen burning minimum mass,
$\sim$0.075~{\Msun} \citep{Burrows1997}, and the onset of molecular lines in the
photosphere, $\sim$0.8~{\Msun} \citep{Kirkpatrick1991}---has limited
studies to small, nearby samples. Almost two decades ago, the
Palomar--Michigan State University (PMSU) survey cataloged the spectra
for $\sim$3,000 M dwarfs (dMs) and was the largest such study \citep{Reid1995,Hawley1996}.
Other samples used to study the distribution of low-mass binaries and to
calibrate low-mass stellar properties using binaries were even smaller,
with 30--100 systems \citep{Fischer1992,Henry1993,Reid1997,Delfosse2004}.  
Hence, these studies often lacked the statistical robustness for firm,
independent results, which were then often tethered to and studied in
comparison with higher-mass stars. In addition, modeling efforts for the
low-mass late-K and M dwarfs have been stymied by incomplete molecular
line lists resulting in uncertain opacities caused by the molecules in
the stellar photosphere at effective temperatures of $\lesssim$4300~K
\citep*{Hauschildt1999a}. As a result of these observational and
modeling challenges, low-mass stars have not been well characterized,
and our techniques for measuring their properties are ill-defined.
However, low-mass stars comprise of $\gtrsim$70\% of the
stars in the Galaxy \citep{Henry1998} and are the best tracers of its
distribution and (at least nearby) structure
\citep[e.g.][]{Bochanski2010}. Low-mass stars also have 
lifetimes longer than that of the Galaxy \citep{Laughlin1997}, making
them the ideal tracers of its formation, chemical, and dynamical history.

The advent of deep, all-sky surveys has revolutionized low-mass star
and brown dwarf science. In particular, the Sloan Digital Sky Survey
\citep[SDSS;][]{York2000}, the Two Micron All Sky Survey
\citep[2MASS;][]{Cutri2003}, and UKIRT Infrared Deep Sky Survey
\citep[UKIDSS;][]{Lawrence2007}, and Wide-field Infrared Survey
Explorer \citep[e.g.,][]{Wright2010} have been critical in expanding sample
sizes. SDSS alone has enabled a photometric catalog of $>$30 million 
\citep{Bochanski2010} and spectroscopic catalog of $>$70,000
\citep[][hereafter, W11]{West2011} low-mass stars. \citet{Munn2004,Munn2008} combined
the USNO-B and SDSS astrometry to calculate proper motions for SDSS
photometric catalogs. This proper motion catalog is 90\% complete down
to $g < 19.7$, with typical errors of 5--7{\masyr}. However, the lower
resolution limits the catalog to sources separated by $\sim 7\arcsec$.

\begin{figure*}[Htb]
  \begin{centering}
  \includegraphics[width=1\linewidth]{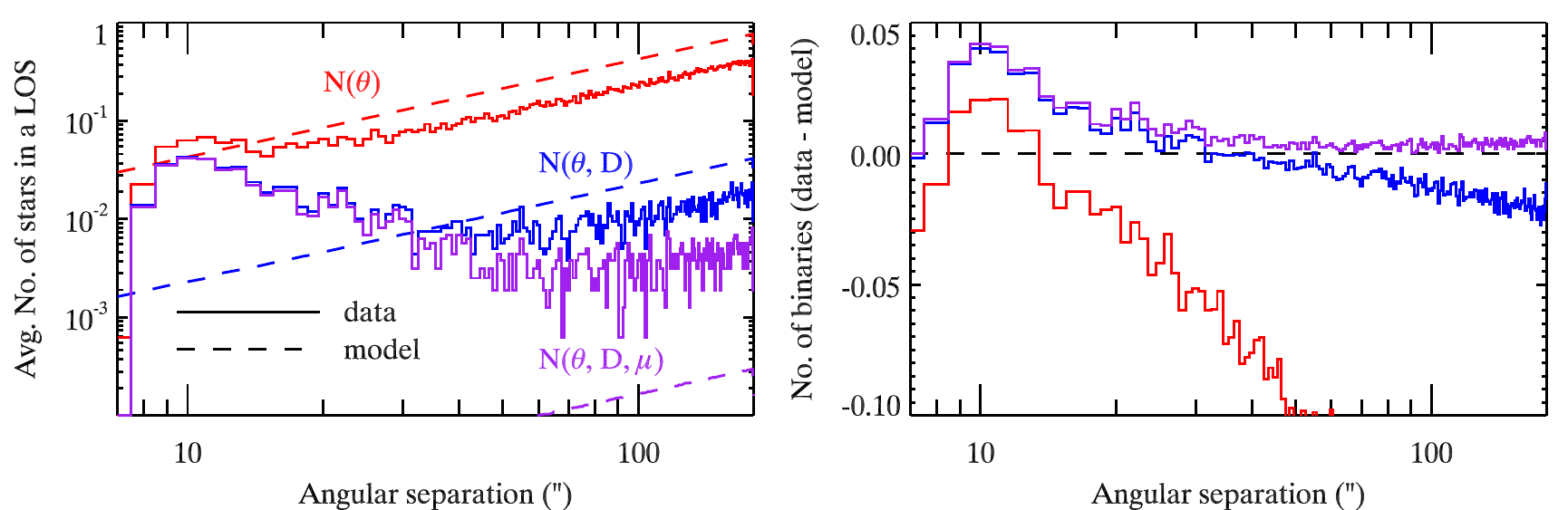}
  \caption{The average number of visual companions (left) and real
    binaries (right), as assessed by our Galactic models, as a function of angular
    separation in $>$1600 lines-of-sight where binary candidates were
    identified in the {\slowpokes} catalog \citepalias{Dhital2010}.
    The solid histograms shows the SDSS distributions while the dashed
    histograms show the distributions generated in our Monte Carlo
    models. In this figure, we show the distribution of the average number of companions
    around $\gtrsim$1600 candidates. We succesively match the angular separation
    (with the results shown in red histograms), photometric distances
    (blue), and proper motions (purple). The model distributions predict more stars at
    larger angular separations as they include all stars whereas the
    data count only the stars that are detected in SDSS.
    Intriguingly, the blue and purple histograms
    were almost identical up to 20--25$\arcsec$, suggesting proper
    motions were not needed to identify those binaries. This is seen
    even more clearly in the right panel. In this paper,
    we conduct a search for binary companions around $\sim$24 million
    low-mass stellar sources in the SDSS survey without using proper
    motions.
  }
  \label{Fig:slowpokes}
  \end{centering}
\end{figure*}

\citet[][hereafter, \citetalias{Dhital2010}]{Dhital2010} identified
the Sloan Low-mass Wide Pairs of Kinematically Equivalent
Stars ({\slowpokes}) catalog of wide binaries by matching positions, photometric
distances, and vector proper motions. The fidelity of each pair was
assessed by a six-dimensional Galactic model, which was built using
empirical stellar number density \citep{Juric2008,Bochanski2010} and
space velocity distributions \citep{Bochanski2007a}. With 1342 common
proper motion pairs, with $\leq$5\% probability of chance alignments,
{\slowpokes} was the largest catalog of low-mass, wide binaries and
contained a diverse set of pairs with G/K$+$dM, white dwarf (WD)$+$dM,
dM$+$dM, and M subdwarf (sdM$+$sdM) systems. {\slowpokes} has enabled a variety of
follow-up observations to probe higher-order multiplicity \citep{Law2010},
metallicity \citep{Dhital2012}, activity \citep{Gunning2014}, rotation,
and the age--activity relation (Morgan et al., {\it in preparation},
Massey et al., {\it in preparation}) of M
dwarfs. In addition, \citep{Andrews2012,Andrews2014} have adapted the technique
used for {\slowpokes} to identify double the sample of known wide
WD$+$WD binaries.

The {\slowpokes} catalog was restricted at small angular separations
($\theta \leq 7\arcsec$) and at lower masses ($\lesssim$~M6) by its
dependence on the SDSS/USNO-B proper motions 
\citep{Munn2004}. Specifically, due to the shallow faintness limit of
USNO-B, a significant fraction of mid--late dMs that are detected in
SDSS do not have USNO-B counterparts and, therefore, proper motions. Similarly, the
resolution of USNO-B sets a separation limit of $\sim 7\arcsec$ for the
SDSS/USNO-B proper motion catalog. These incompleteness are
inherited by the {\slowpokes} catalog. In addition, using proper
motions to identify binaries precludes systems that are either at
large distances or nearby but moving slowly with respect to the Local
Standard of Rest. For example, we used a minimum proper motion of
40~{\masyr} for the {\slowpokes} catalog.

Results from the Galactic model indicated that visual binaries with small angular
separations could be identified at a high level of fidelity by
matching photometric distances alone \citepalias{Dhital2010}. 
Figure~\ref{Fig:slowpokes} (left) shows Figure~5 from
\citetalias{Dhital2010}. The solid red histogram shows the number of stars,
averaged among $gtrsim$1600 lines-of-sight where binary candidates were
identified, as a function of angular separation, as measured in the
SDSS DR7 photometric catalog. The dashed red histogram shows the
distribution for single, non-associated stars in the same fields  as
simulated by our Galactic model. In essence, this was a measure of the
likelihood of whether the companion stars were real binaries or simply
chance alignments. The blue and purple histograms show the
distribution when the photometric distances and proper motions were
matched. Each addition of a dimension resulted in a
rejection of a large number of chance optical pairs. The larger stellar number
counts at large separations in our simulations (as compared to the data) was
caused largely by the fact that the data were limited to $r \lesssim
20.5$ as the SDSS/USNO-B proper motions were required whereas our
Galactic model simulates \textit{all} stellar objects.
Two features stood out in this figure: (1) the excess of pairs at small
separations, which was how \citet{Michell1767} first identified binary
systems, or ``double'' stars as he called them and (2) the blue and
purple histograms were essentially the same until about $\theta \sim
20-25\arcsec$, suggesting those binaries could have been identified
without proper motions. This is even more evident in the right panel
of Figure~\ref{Fig:slowpokes}, where we have plotted the difference
between the data and model distributions. These figures clearly
demonstrate that proper motions are not required to identify binaries
up to a critical separations with a high level of fidelity. A caveat
is that the critical angular separation obviously depends on the
magnitude limit of the sample. For the {\slowpokes} sample with a
limiting magnitude of $r=20.5$ and proper motion $>$40~{\masyr},
$>$90\% of the binary candidates within 20$\arcsec$  with matching photometric distances
also had matching proper motions and were classified as CPM pairs.

In this paper, we extend the {\slowpokes} catalog by identifying
binary systems with angular separations of 1--20$\arcsec$ based
entirely on SDSS photometry and astrometry. This allows us to identify
visual binaries to $r=22.2$, with significant numbers at the mid--late
M spectral types. In Section~\ref{Sec:data} we describe the
initial sample of low-mass stars that we search around. As in
\citetalias{Dhital2010}, our search algorithm is based on matching
angular separation and photometric distances supplemented by a Monte
Carlo-based Galactic model, which is described in Section~\ref{Sec:algorithm}.
We discuss the characteristics of the resultant binary sample in
Section~\ref{Sec:results}. We examine the currently debated formation
theories for wide stellar binaries and VLM/BDs in light of {\slowpokes}-II
sample of binaries in Section~\ref{Sec:discussion}. We summarize our
results in Section~\ref{Sec:conclusions}.

The {\slowpokes} and {\slowpokes}-II catalogs, along with followup
spectra, are publicly available
online.\footnote{\label{fnote1}\url{http://slowpokes.vanderbilt.edu}}. 

\section{SDSS Data}\label{Sec:data}
One of the largest and most influential astronomical surveys ever
conducted to date, SDSS is a comprehensive imaging and spectroscopic
survey \citep{York2000}. Over eight years of operation between
2000-2008, it collected imaging and spectroscopic data for over a decade
years using a dedicated 2.5-m telescope at Apache
Point Observatory, New Mexico \citep{Gunn2006}. The telescope has
a 120 megapixel camera that has a field of view of 1.5
square degrees \citep{Gunn1998} and conducts imaging in five broad 
optical bands ($ugriz$) between $\sim$3,000 and 10,000~$\AA$
\citep{Fukugita1996}. The last data release with imaging data, Data
Release 8 (DR8), comprised of $\sim$450~million unique objects over 14,555 square
degrees of the sky, spanning the entire northern sky as well as the
Southern Galactic Cap \citep{Aihara2011}. The global
absolute astrometric precision was 70~mas \citep{Pier2003}
while the photometry has relative calibration accuracies of 2\% in the
$u$ band and 1\% in the $griz$ bands \citep{Padmanabhan2008}. The
catalog is 95\% complete for point sources of 22.0, 22.2, 22.2, 21.3,
and 20.5 in the $ugriz$ bands, respectively \citep{Gunn1998}.
The corresponding spectroscopic survey has obtained  $\sim$1.8 million
optical spectra, with $\lambda/\Delta\lambda \approx 2000$, over 9274
square degrees \citep{Aihara2011} using a pair of fiber-fed double
spectrographs. The Third Sloan Digital Sky Survey
\citep[SDSS-III;][]{Eisenstein2011}, which is comprised of four
different spectroscopic surveys, is currently underway.

The DR8 photometric catalog has more than 200 million point
sources. As in \citetalias{Dhital2010}, we used the Catalog Archive 
Server query tool (CasJobs\footnote{\url{http://skyserver.sdss3.org/CasJobs/}})
to select the sample of low-mass stars from the DR8 \textsc{star}
table as having  $r-i \geq 0.3$ and $i-z \geq 0.2$, consistent with
spectral types of K5 or later \citep{West2008}. Selecting from the
\textsc{star} ensures that the object is a \textsc{primary} and not a
duplicate detection and that its morphology is consistent with being a
point source. This morphological classification is $>$95\% accurate up
to $r\approx 21.5$ \citep{Lupton2001}.  Even up to $r\approx 22.5$,
stars outnumber galaxies $>$8:1 \citep{Fadely2012}, and, therefore, we do
not expect compact galaxies as frequent interlopers in our sample. 
To ensure excellent photometry, we made cuts on the standard quality
flags---all of \textsc{peakcenter, notchecked, psf\_flux\_interp,
  interp\_center, bad\_counts\_error, saturated} were set to be
0 (\citealt{Bochanski2010}; \citetalias{Dhital2010})---and required
the errors in PSF magnitudes to be $\leq$~0.10. Throughout our
analysis, these quality cuts were performed only for the bands that
were used in the analysis for that particular star: $iz$ for M7 or
later stars, $riz$ for K5--M7 stars, $griz$ for F0--K5 stars, and
$ugriz$ for white dwarfs. While no faintness limits were specifically
adopted, the requirement on the error in PSF magnitudes effectively
limits resulting samples to the SDSS 95\% completeness limits.

The resulting sample yielded 33,589,670 stellar sources, with colors
consistent with K5--M9 dwarfs. In \citet{Dhital2010} we found that
chance alignments were unacceptably high near the Galactic Plane due to
the higher stellar density and the higher uncertainties caused by
higher extinction. Therefore, we rejected stars at Galactic
latitudes less than 20$\degr$. We also rejected stars with photometric
distances larger than 2500~pc as the distance uncertainties are larger than
$\sim$350~pc, making distance matching meaningless at those
distances (see Section~\ref{Sec:dist_ms} for further discussion). As a
result, we started with an initial sample of 24,036,982 low-mass
stars., around which to search for companions.

\section{Method: Identifying Binary Candidates}\label{Sec:algorithm}

\subsection{Assessing the SDSS source classification algorithm}\label{Sec: completeness}
The tighest binary we were able to identify was $\gtrsim$1$\arcsec$ (see
Figure~\ref{Fig:sep} below), suggesting a fundamental limitation in
our technique at that value. However, with a plate scale of
0$\farcs$396~pixel$^{-1}$, SDSS should be able to distinguish tighter
binaries. Both unresolved and partially-resolved binaries, with
separations smaller than the resolution limit or the plate scale, have
been identified from both the 2MASS \citep{Kraus2007a} and the Palomar
Transient Factory \citep{Terziev2013} photometric catalogs, albeit
aided by multi-epoch data in the latter case. Therefore, we conducted
an examination of (1) what was restricting us from identifying
binaries at $<$1$\arcsec$ and (2) what fraction of resolved binaries
we were missing in our resultant sample due to this limitation.

First, we briefly describe the SDSS point source classification
scheme. The algorithm is fully documented on the SDSS
webpages\footnote{\url{http://www.sdss3.org/dr9/algorithms/}} and
is beyond the scope of this paper, thus we only outline the scheme
that classifies identified sources into the \textsc{STAR} table. The
SDSS pipeline uses \textsc{Resolve} to identify unique sources and
and \textsc{DEBLEND} to deblend them when multiple sources are present. All of the
identified sources, for which photometric parameters have been measured, are
cataloged in the \textsc{PhotoObjAll} table. This table has
several sub-tables, known as \textit{view}s in SDSS:
\textsc{PhotoObj}, which includes all primary and secondary objects;
\textsc{PhotoPrimary}, which includes all the primary detections or
ones classified as the best version of the objects; \textsc{PhotoSecondary}, which
includes the duplicate detection(s); and \textsc{PhotoFamily}, which
includes the original undeblended source as well as the sources for
which the deblending failed.  Based on the morphology, the
\textsc{PhotoPrimary} is further divided into \textsc{Star} (point
sources), \textsc{Galaxy} (extended sources), \textsc{Sky} (sky
samples), and \textsc{Unknown} (unclassified sources). 
A \textsc{PhotoPrimary} object is classified as a
point source and included in \textsc{Star} when a PSF fit provides a
good approximation to its light profile; otherwise, it is classified
as an extended source\footnote{More specifically, a point source has
  \textsc{psfMag} $-$ \textsc{cmodelMag} $\leq$ 0.145, where \textsc{psfMag} and
  \textsc{cmodel} are the magnitudes measured by fitting a PSF model
  and a linear combination of de Vaucouleurs and exponential models,
  respectively, for an object's light profile.}. 

As the table with all the unique point source objects, \textsc{Star}
is the repository used by all stellar studies that use the SDSS survey
\citep[e.g.,][]{Covey2007,Juric2008,Bochanski2010} including this
study. Therefore, it is troublesome that we detected no resolved
companions within 1$\arcsec$ in our search. Given the scope of \textsc{Star},
it is of utmost importance to understand its completeness level.
In particular, it is necessary to quantify the sources
that were detected in SDSS but are missing from \textsc{Star}, i.e.,
the sources that were misclassified. This would allow any individual
study to apply relevant corrections for the sources that were truly
missing from SDSS survey (e.g., faintness limit, unresolved binarity).

To test the completeness of \textsc{Star}, a sample of bona fide
stellar sources needs to be used. Fortunately, there exist large spectroscopic
samples of confirmed stellar sources in the SDSS catalog. We chose the
DR7 catalog of 70,841 M dwarfs from the SDSS \cite{West2011}. Each
spectrum in this catalog was inspected by eye and verified to be a
bona fide M dwarf with relatively high signal-to-noise. Therefore,
looking at how these bona fide M dwarfs are cataloged in \textsc{Star}
using only photometric information and how they are flagged should
help us understand why we were unable to find binaries tighter than
1$\arcsec$. As the selection algorithm for M dwarfs in the SDSS
spectroscopic survey was not based off \textsc{Star} \citep{West2011},
this is an ideal sample to investigate how many of the 70,841 bona
fide M dwarfs are not included in \textsc{Star}. 

\subsubsection{What is the completeness of DR7 M dwarf sample?}
We performed a \textsc{plate}/\textsc{mjd}/\textsc{fiberID}-based
search for the 70,841 M dwarfs in DR7 \textsc{STAR} using the SDSS
CasJobs portal and recovered only 66,001 ($\sim$93.2\%). When the same
search was performed on DR7 \textsc{PhotoObj}, every single
spectroscopic target had an counterpart \footnote{This search was done
  in DR7, from which the spectroscopic catalog was 
  compiled. As each DR is completely new reduction, it is expected
  that not 100\% would be recovered between different DRs. Indeed, we
  recovered fewer objects when we performed the same search in DR8 and
  DR9 catalogs.}. 
Appropriate flag cuts (see Section~\ref{Sec:data}) were used in both
searches; when excellent photometry was not present, we assumed the
sources were not real. That every single M dwarf was in
\textsc{PhotoObj} but $\sim$7\% were missing in \textsc{Star} made it
evident that they had been misclassified.

The first type of misclassification (544
instances; $\sim$0.77\% of all sources) happened when an M
dwarf was imaged two or more times, and the \textsc{Primary} detection 
failed the photometric flag cuts but one or more \textsc{Secondary}
detection(s) passed them. About one-third of the SDSS footprint was
imaged multiple times, mostly when plates overlapped each other. In
such cases the \textsc{Primary} / \textsc{Secondary} classification is
based on  whether the field was designated as the ``primary'' field
for that area of the sky. As the quality of photometry plays no part
in this classification, this is understandable. In fact, it is
remarkable that only 0.77\% of sources are lost; this incompleteness
is largely negligible.

The second kind of misclassification, which happened in 4,295
instances, happened when the M dwarfs were classified as extended
sources. This was either because of an nearby extended source
contaminated and extended the light profile of the M dwarf or there
were two partially resolved stellar sources. (The latter are exactly
the kind of sources we are looking for in our binary search.) That
$\sim$6\% of all stars could be not cataloged in \textsc{Star} is
rather alarming. We note that this effect is likely exaggerated for
the spectroscopic sample (as compared to the entire photometric
sample) as a large number of M dwarf spectra were acquired as they
were targeted as potential quasars and LRGs
\citep{Adelman-McCarthy2006}. Thus, they are
more likely to be flagged as being extended sources.  However, an
incompleteness of $\sim$6.1\% is significant and should be properly
accounted for studies that use the SDSS photometric sample.

\subsection{Photometric Distances}\label{Sec:dist_ms}
\begin{center}
  \begin{deluxetable*}{llll}
    \tablecolumns{21}
    \tabletypesize{\scriptsize}
    \tablewidth{0pt}
    \tablecaption{Photometric parallax relations used in this paper}
    \tablehead{
      \colhead{type} &
      \colhead{locus} &
      \colhead{Photometric parallax relation} &
      \colhead{References}
    }
        \startdata
    F0--K4       & $-$0.01 $\le$ (r$-$z) $<$ 0.50 & M$_g$ $=$ 2.845 $+$ 1.656~(g$-$i) $+$ 3.863~(g$-$i)$^2$ $-$ 1.795~(g$-$i)$^3$    & \citet{Covey2007} \\
    K5--M9       & 0.50 $\le$ (r$-$z) $<$ 4.53    & M$_r$ $=$ 5.190 $+$ 2.474~(r$-$z) $+$ 0.4340~(r$-$z)$^2$ $-$ 0.08600~(r$-$z)$^3$ & \citet{Bochanski2010} \\
    L0--L9       & 1.70 $\le$ (i$-$z) $\le$ 3.20  & M$_i$ $=$ $-$23.27 $+$ 38.40~(i$-$z) $-$ 11.11~(i$-$z)$^2$ $+$ 1.064~(i$-$z)$^3$    & \citet{Schmidt2010} \\
    M subdwarf   &  (g$-$r) $>$ 1.50; 0.8 $<$ r$-$z $<$ 2.5 & M$_r$ $=$ 7.9547 $+$ 1.8102~(r$-$z) $-$ 0.17347~(r$-$z)$^2$ $+$ 7.7038~$\delta_{{g-r}}$ $-$ 1.4170~(r$-$z)~$\delta_{{g-r}}$ & \citet{Bochanski2013} \\
    white dwarf & \citet{Girven2011} & iterative fits to DA cooling models & \citet{Harris2006} \\
    \enddata
    \label{Tab:plx}
  \end{deluxetable*}
\end{center}
\begin{figure}[Bht]
  \begin{centering}
  \includegraphics[width=1\linewidth]{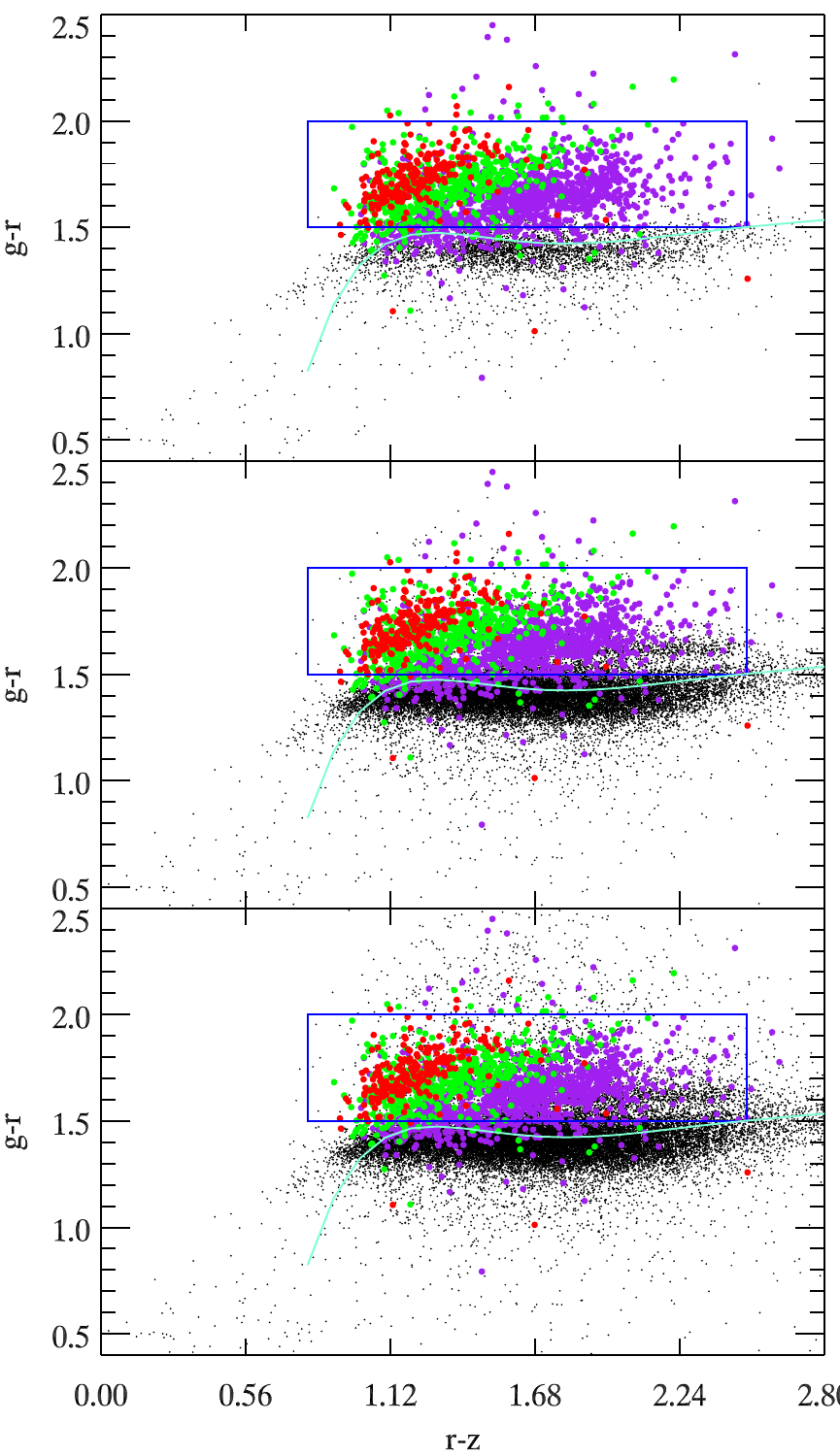}
  \caption{The M subdwarfs, extreme subdwarfs, and ultra subdwarfs
    (purple, green, and red circles, respectively) are clearly
    separated from the M dwarfs (black dots). The separation is even
    cleaner when the photometry is precise; the bottom panel shows all
    the M dwarfs in SDSS DR7 catalog \citep{West2011} while the middle
    and top panels only show ones with uncertainties $\leq$0.05 and
    0.02 mags, respectively. The blue box shows the subdwarf locus
    identified in \citet{Bochanski2013} for which photometric
    distances can be calculated.
  }
  \label{Fig:sdm}
  \end{centering}
\end{figure}
\vspace{-18pt}
\noindent\textbf{Main-sequence Dwarfs:} 
We determined distances to main-sequence dwarfs using photometric
parallax relations that are calibrated from directly-measured
trigonometric parallaxes. Even though there is no spectroscopic
confirmation, MS dwarfs dominate the stellar sources in the SDSS
catalog \citep{Covey2007}. Three separate photometric parallax relations
were used, such that the chosen magnitude and colors trace the stellar
temperature monotonically in that regime:  M$_g$ vs. $g-i$ for F0--K4
\citep{Covey2007}, M$_r$ vs. $r-z$ for K5--M9, \citep{Bochanski2010},
and M$_i$ vs. $i-z$ for L0--L9 \citep{Schmidt2010} dwarfs. The 
photometric parallax relations are given in Table~\ref{Tab:plx}.

Magnitudes were corrected for extinction, with values from \citep*{Schlegel1998} as
tabulated in the SDSS database, were used in all cases. We adopted an
error of 0.3~mag in the calculated absolute magnitudes, which implies 
a 1$\sigma$ error of 14\% in distance. This scatter is caused by a
combination of metallicity, magnetic activity, and unresolved
binarity \citep{West2005,Sesar2008,Bochanski2011}. The presence of magnetic activity and higher metallicity
increases the the intrinsic brightness of a star by as much as a magnitude
\citet{Bochanski2011}.  An unresolved binary companion increases the
apparent magnitude by as much as 0.75~mag. Based on high-resolution imaging
studies, components of wide binaries are highly likely to harbor close
companions \citep{Law2010}. Therefore, we are selecting against triple
systems that contain an unresolved binary. However, none of these
parameters can be measured or even estimated from the photometry alone.
\citet{Ivezic2008b} used a photometric metallicity in their
photometric parallax relation, but that is applicable only to FGK
dwarfs. No such relation is known for M dwarfs, except when they are
significantly metal-poor subdwarfs (see below). Therefore, we chose to
ignore the metallicity-dependence.

\vspace{10pt}
\noindent\textbf{M Subdwarfs:}
M subdwarfs are low-metallicity, main-sequence dwarfs typically
associated with the thick disk or the halo. In the M spectral type, the subdwarfs can
be differentiated from the dwarfs by the depleted TiO feature in the
optical spectrum \citep{Gizis1997}. In \citetalias{Dhital2010}, we
used the reduced proper motions to select subdwarf candidates
and identify 70 CPM binaries as subdwarf systems. As a photometric
parallax relation had not been calibrated for M subdwarfs, we used the
one for M dwarfs. As a result the distances were overestimated and
we were able identify only a subset of the subdwarf pairs. 
\citet{Bochanski2013} have since developed a statistical parallax
relation using the SDSS DR7 subdwarf catalog \citep*{Savcheva2014}
based on their redder $g-r$ colors. 
Likely caused by increased hydride absorption, the redder $g-r$
color causes the M subdwarf locus to clearly separate from the M dwarf
locus in the ($r-z$, $g-r$) space \citep{West2004,West2011,Lepine2008a}.

Distinguishing the subdwarfs from the dwarfs, however, is not
trivial. As shown in Figure~\ref{Fig:sdm}, while the loci clearly separate in the ($r-z$, $g-r$)
space, the tail of the M dwarf distribution scatters into the subdwarf
locus.  Given their vastly larger number in the Solar neighborhood,
the dwarfs overwhelm the subdwarf population. However, we found that
the scatter is in large part due the uncertainty in the magnitude
measurements. In Figure~\ref{Fig:sdm}, the spectroscopically
confirmed M dwarfs \citep{West2011} are plotted as black dots while
the spectroscopically confirmed subdwarfs are plotted as purple
(subdwarfs), green (extreme subdwarfs), and red (ultra subdwarfs). In the
bottom panel, where all the stars  plotted, the dwarfs scatter
into subdwarf locus \citep[blue box;][]{Bochanski2013}
significantly. However, when only the dwarfs with uncertainties
$\leq$0.05 (middle) and $\leq$0.02 mag (top) in the $griz$ bands are plotted,
the scatter decreases. While the photometric accuracy for SDSS is $\sim$1\% in the
$griz$ bands, imposing a 1\% or 2\% cut seems to exclude a large
number of sources. This also biases the sample against fainter and
more distant stars like the M subdwarfs. However, when available, exquisite photometry, can
be used to select subdwarfs without the need for spectra. 
This could possibly be extremely beneficial to future photometric
surveys like the \textit{Large Synoptic Survey Telescope}.  

We chose to accept photometry with uncertainties $\leq$0.05 mag to
identify subdwarf candidates. We recognized that this will include
some M dwarf interlopers and increase the rate of false
positives. However, given the low number of subdwarf binaries known,
this is an acceptable risk. We did not include the identified subdwarf
systems in the statistical analysis for this paper. We did not include
the subdwarf binaries in the analysis for this paper, as they are
less likely to be bona fide systems compared to the rest of our
sample. 

We calculated photometric distances to subdwarf candidates
that fit the following criteria:
\begin{align}\label{Eq:sdm}
  g-r &> 1.5 \nonumber \\
  0.8 < r-z &< 2.5 \\
  {\rm psfMagErr}_{griz} &<=0.05 \nonumber 
\end{align}
from the \citet{Bochanski2013} relations. The uncertainty in the
absolute magnitude is $\sim$0.41~mag, which translates to a 20\%
uncertainty in the distance. There were over six million subdwarf
candidates, as defined by Eq.~\ref{Eq:sdm}, in the SDSS photometric
catalog.

\vspace{10pt}
\noindent\textbf{White Dwarfs:}
In \citetalias{Dhital2010} we used proper motions to calculate the
reduced proper motions and identify potential WD candidates and search
for WD companions to the low-mass star sample. Based on $ugriz$
photometry alone, there is no conclusive way to identify potential WD
companions around our low-mass dwarf sample. However, in the
color--color diagrams, WDs segregate from the main-sequence 
stars and the quasars. In particular, \citet{Girven2011} have defined 
a ($g-r$, $u-g$) locus for hydrogen-atmosphere WDs (DAs) based on a
sample of spectroscopically identified DAs \citep{Eisenstein2006} in
SDSS DR7. The efficiency of the photometric selection for the
spectroscopic sample was only $\sim$62.3\%, after non-DA WDs including
WD$+$MS pairs ($\sim$8.9\%), early-type MS stars and subdwarfs
($\sim$11.3\%), and quasars ($\sim$17.2\%) were
removed. However, as quasars were specifically targeted for the SDSS
spectroscopic sample, they are disproportionately represented. In the
larger photometric sample, the contamination due to quasars is likely
to be minor \citep{Girven2011}. Therefore, we use
the \citet{Girven2011} ($g-r$, $u-g$) locus to identify potential WDs
but note that $>$20\% of the WD candidates will be interlopers and
will require spectroscopic confirmation.

We calculated the photometric distances to candidate DAs by fitting
the  $ugriz$ photometry to WD cooling models \citep{Bergeron1995}, as
specified in \citet{Harris2006}. This algorithm fits the photometry to the
model in an iterative manner to derive a bolometric luminosity and a
distance. However, the composition and mass/gravity is degenerate when
only the photometry is available and cannot be determined. So we used
the models with pure hydrogen atmospheres and a gravity of $\log~g=8.0$.
As our adopted ($g-r$, $u-g$) locus selected only DAs, the composition
should introduce significant uncertainties in the distances. However, 
incorrect distances will be derived for WDs with unusually high
mass/gravity ($\sim$15\% of all WDs), unusually low mass/gravity
($\sim$10\% of all WDs), or helium-dominated atmospheres
\citep{Harris2006}. 

A comparison of our photometric distances and the spectroscopic
distances for the DA WDs in the SDSS DR7 catalog
\citep{Kleinman2013} showed a 14--20\% scatter (J. Andrews,
\textit{private communication}). To be conservative in our matching
process, we adopted 14\% as the error in our photometric distances.
There were 56,505 WD candidates, as selected using the
\cite{Girven2011} ($g-r$, $u-g$) locus, in the SDSS photometric
catalog.

\subsection{Binary Candidate Selection}\label{Sec:candidate_match}
We searched for stellar sources that had been classified as
\textsc{Primary} detections within angular separations of
1--20$\arcsec$ of our sample of 24,036,982 low-mass stars. The inner
search radius of 1$\arcsec$ was determined by the SDSS database, as
discussed in Section~\ref{Sec: completeness}. We chose the outer
search radius to be 20$\arcsec$, after which the number of chance
alignments for pairs without proper motions were significant in
\citetalias{Dhital2010} (see Figure~\ref{Fig:slowpokes}). The search was conducted
using in the \textsc{Neighbors} table in the SDSS CasJobs Query interface.
Photometric quality cuts, as described above, were performed. We then
matched the photometric distances ($d$) to within 1$\sigma$. However, as the
error in the distances is a percentage error, we also required the
difference in the distance to be less than 100~pc to be classified as
a candidate binary pair. Thus, our binary candidate pairs matched the
following criteria:
\begin{align}\label{Eq:pair_criteria}
\theta &= 1-20\arcsec \nonumber\\
\Delta\ d &\le {\rm min}(1\sigma_{\Delta~d}, 100~{\rm pc})\nonumber\\
d &\le 2500~{\rm pc}\\
|b| &\ge 20\degr.\nonumber
\end{align}
All stars were selected to have good photometry and \textsc{psfMagErr} $\le
0.10$~mag for the bands that are used  in their selection and
analysis. We identified 514,424 dM$+$MS, 1212 sdM$+$sdM, and 642
WD$+$dM candidate pairs. 

\subsection{The Galactic Model: Assessing False Positives in the
  Binary Candidate Sample}\label{Sec:model}
\begin{center}
\begin{deluxetable}{lllc}
\tablewidth{0pt}
\tablecaption{Galactic Structure Parameters}
\tablehead{
  \colhead{Component}    & \colhead{Parameter name} & 
  \colhead{Parameter description} & \colhead{Adopted Value}}
\startdata
           & $\rho\ (R_{\odot}, 0)$ & stellar density & 0.0064 \\
\hline
           & $f_{\rm thin}$  & fraction\tablenotemark{a} & 1-$f_{\rm thick}$-$f_{\rm halo}$ \\
thin disk  & $H_{\rm thin}$  & scale height     & 260~pc \\ 
           & $L_{\rm thin}$  & scale length     & 2500~pc \\
\hline
           & $f_{\rm thick}$ & fraction\tablenotemark{a} & 9\% \\
thick disk & $H_{\rm thick}$ & scale height     & 900~pc \\
           & $L_{\rm thick}$ & scale length     & 3500~pc \\
\hline
           & $f_{\rm halo}$  & fraction\tablenotemark{a} & 0.25\% \\
halo       & $r_{\rm halo}$  & density gradient & 2.77 \\
           & $q\ (=c/a)\tablenotemark{b}$    & flattening parameter & 0.64 \\            
\enddata
\label{Tab:gal_model}
\tablenotetext{a}{Evaluated in the solar neighborhood}
\tablenotetext{b}{Assuming a bi-axial ellipsoid with axes {\em a} and {\em c}}
\tablecomments{The parameters were measured using M dwarfs for the
  disk \citep{Bochanski2010} and main-sequence turn-off stars for the
  halo \citep{Juric2008} in the SDSS footprint.}
\end{deluxetable}
\end{center}
Despite the rigorous nature of our candidate selection, chance
alignments will be present in a sample of wide visual binaries. Such
chance alignments arise from the measurement uncertainties in the
parameters used in the selection criteria, as well as via the 
the inherent spreads in these parameters in the Galaxy
\citepalias{Dhital2010}. The number of chance alignments grows as a 
function of the angular separation ($\propto \theta^2$) and
distance. As our search does not include any kinematic matching, the
probability of chance alignment is particularly high and, therefore,
needs to be rigorously assessed for each and every candidate binary
pair. Such a quantitative assessment sifts out false
positives from the sample. In \citetalias{Dhital2010} we built a
Monte Carlo-based Galactic model that recreated the stellar
populations along the line-of-sight (LOS) of a candidate binary and
calculated the probability of chance alignments. Based on
empirically-measured parameters for the Milky Way, the model accounted
for the variations in the stellar number density and space velocities, which
become important beyond the Solar Neighborhood.
We employed the same model to assess the probability of chance
alignment for the binary candidates identified here. The model was
described in detail in \citetalias{Dhital2010}; here we only provide a
brief synopsis.

Instead of the computationally implausible task of simulating the entire
Galaxy with $\gtrsim 10^{11}$ stars, we recreated a $30\arcsec \times
30\arcsec$ cone centered at the ($\alpha$, $\delta$) of each primary out
to a distance of 2500~pc. First, we calculated the total number of
stars in the conical volume by integrating stellar number density
profiles that assume a bimodal disk \citep{Bochanski2010} and an
ellipsoidal halo \citep{Juric2008}. The model parameters are given in
Table~\ref{Tab:gal_model}. While a two-component model is an
oversimplification of the complicated scale height distribution
\citep*{Bovy2012}, it is easy to model and suffices for our purpose
of recreating a random Galaxy. Each LOS was then repopulated with
stars using the rejection method \citep{Press1992}. The rejection
method ensured that the stars were randomly redistributed while
following the overlying stellar number density distribution
function. Each star that is generated has a $\alpha$,  
$\delta$, and distance\footnote{In \citetalias{Dhital2010} we modeled the
3D velocities for each star, but we do not do so here
as we are not utilizing proper motions}. The total
number of stars in an LOS ranged between 2000--10000, enough to both
produce small-scale density variations as are seen in the Milky Way
and perform calculations with relative ease. However, as the number 
density profiles are smoothed functions that were made to fit to the
Galactic field, our model does not produce larger density variations
like moving groups, open clusters, or streams.

With the simulated galaxy, we then counted how many stars were found
in the ellipsoid that is centered at the  $\alpha$ and $\delta$ of the
binary candidate and defined by its angular separation and its
distance errors. This was the same criteria that we used in
the search for candidate systems in the SDSS data. Indeed, to conduct
the exact same search, we chose to not convolve the luminosity or mass
function into the model and rather searched for all stars. As all stars in our
simulation are single stars, any star that satisfied the search
criteria is a chance alignment. Therefore, the average number of
chance alignments is an assessment of the probability that our
candidate binary is a false positive. For each candidate pair, we ran
1000 Monte Carlo realizations. If the probability of chance alignment,
P$_{\rm f} \le 0.10$, we ran a further 4000 realizations for better resolution.

\begin{figure*}[ht]
  \begin{centering}
    \includegraphics[width=1\linewidth]{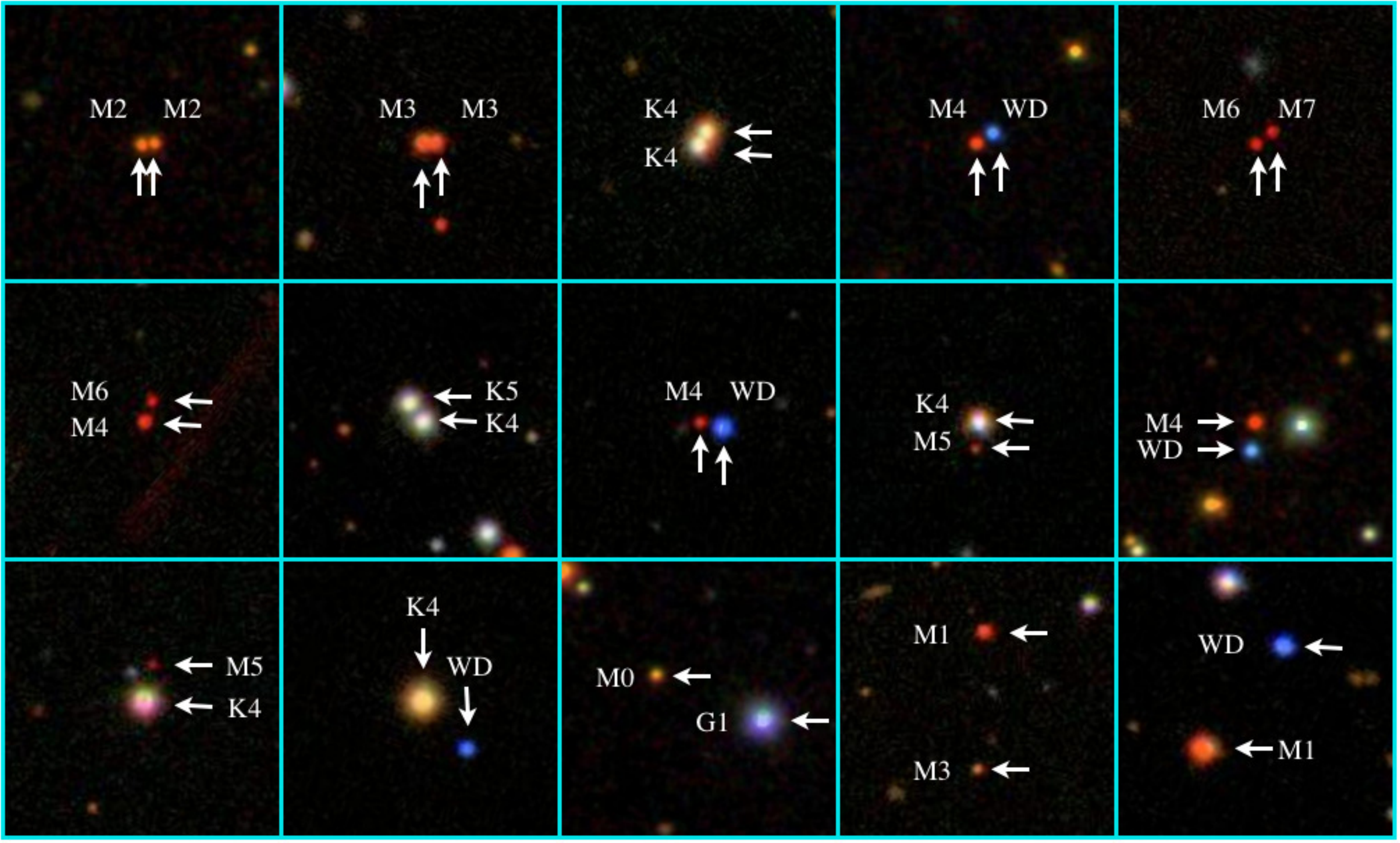}
    \caption{\protect\small
    A $gri$ composite collage of {\slowpokes}-II binaries, 48$\arcsec$
    on a side. We have identified 105,537 \textit{bona fide} binaries,
    with at least one component later than K5, without using proper motions.
    }
    \label{Fig:collage}
  \end{centering}
\end{figure*}

\section{Results: the {\slowpokes}-II Binary Sample}\label{Sec:results}

Following \citetalias{Dhital2010} we classified candidate pairs with a
probability of chance alignment, P$_{\rm f} \le 0.05$ as real
binaries. We note that this limit does not have any physical
motivation but was chosen to minimize the number of spurious
pairs. This cut results in 105,537 dM$+$MS, 450
WD$+$dM, and 944 sdM$+$sdM binary systems with separations of
1--20$\arcsec$. 141 of dM$+$MS binaries are VLM binary candidates
Table~4, with $i-z$ colors redder than the median M7
dwarf for both components. Despite the lack of kinematic information,
we dub this catalog {\slowpokes}-II. The data for the pairs are
tabulated in  and is also available in an online visualization
portal$^5$.

This represents a significant increase over the {\slowpokes} catalog
of 1342 CPM binaries we presented in \citetalias{Dhital2010}. Each
binary has a low probability of chance alignment, with the 
threshold set at {\Pf} $<$ 5\%. Using that threshold, we might expect
5277 of the pairs to be false positives. However, each binary has its
own probability of chance alignment, based on the size of the binary
and its position in the Galaxy, that in most cases is significantly
lower than 5\%. Based on those probabilities, we
expect only 2464 (or 2.33\%) of the binaries to be false positives.

Figure~\ref{Fig:collage} shows a collage of $gri$ composite images,
48$\arcsec$ on a side, of 15 {\slowpokes}-II binaries. The inferred
spectral types of MS dwarfs, based on their $r-z$ colors, are shown
for each component.
Table~3 summarizes the {\slowpokes}-II catalog, with the
properties of the systems and positions and photometry of both
components included. Table~4, Table~5, and
Table~6 summarize the candidate VLM, WD$+$dM, sdM$+$sdM
systems, respectively. While these systems are also analyzed with the Galactic model
and required to have {\Pf} $<$ 5\%, their classification as an sdM,
VLM, or WD is based on photometry alone. This adds a level of
uncertainty that the Galactic model cannot quantify. For that reason,
these samples are likely to be contaminated at a higher rate than the
{\slowpokes}-II sample in Table~3. Thus, we do not
include samples in the analysis presented in this paper.

Both the {\slowpokes} and {\slowpokes}-II catalogs are publicly
available on the worldwide web$^5$
via the Filtergraph portal \citep{Burger2013}. We hope this
will enable to the entire community to interact with and visualize
this large data set using a dynamic, multi-dimensional plotting applet.
The table view Filtergraph is also an easy medium to select targets
for followup observations.

\begin{figure}[htb]
  \begin{centering}
  \includegraphics[width=1\linewidth]{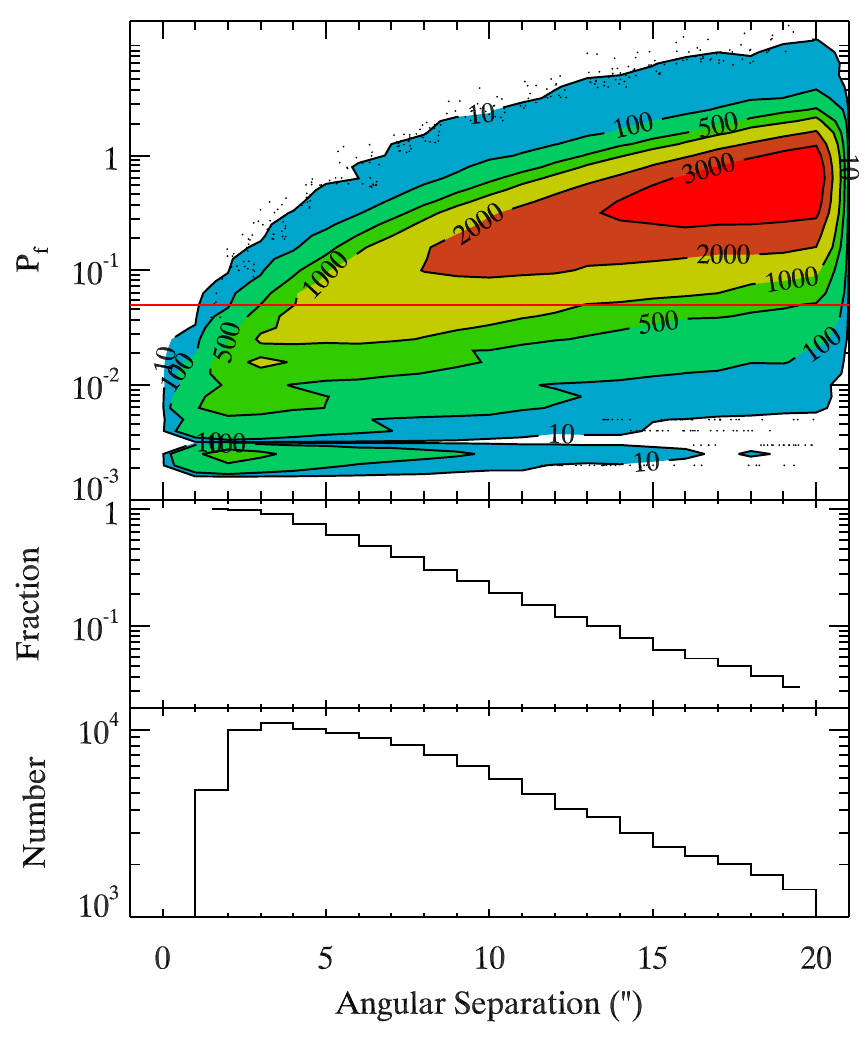}
  \caption{The number density, as represented by the contours, of the
    514,424 binary candidates as a function of {\Pf} as calculated by
    the model vs. the angular 
    separation and distance. The red line is the threshold {\Pf}
    that we adopted for the {\slowpokes}-II catalog. The fraction of
    candidates that passed the {\Pf} threshold---which quantifies the
    number of chance alignments expected within the bounds of that
    particular binary---are shown in the middle
    panels while the distribution of the resultant 105,537 {\slowpokes}-II
    binaries are shown in the bottom panels.
  }
  \label{Fig:pf_sep}
  \end{centering}
\end{figure}

\subsection{Identifying bona fide wide binaries without kinematics}
Figure~\ref{Fig:pf_sep} shows the number density of the 514,424
binary candidates, identified in Section~\ref{Sec:candidate_match}, as a
function of the probability of chance alignment, {\Pf}, and angular
separation. As would be expected, the number of candidate pairs grows
sharply with angular separation. There is also a large spread in {\Pf}
at a given separation, reflecting the varying stellar densities along
different lines-of-sight. Especially at the large angular separations,
the {\Pf} is quite high for a large number of candidates, indicating
that they are chance alignments. Therefore, we applied the {\Pf} $\leq$ 
5\% cut, shown as the red line, for the candidates to be included in the
{\slowpokes}-II catalog. This is a fairly conservative cut but serves
to minimize the number of false positives. It is also the same
threshold that we used in \citetalias{Dhital2010}, although the
{\slowpokes} systems had proper motions and, hence, should have a much
lower rate of false positives. The middle panel shows the histogram of
fraction of candidate binaries that passed this threshold as a
function of angular separation. All of the candidate pairs within
3$\arcsec$ but only $\sim$2\% of the pairs at 20$\arcsec$ were vetted
to be real binaries by the Galactic model. The bottom panel shows the
angular separation histogram for the 105,537 binaries in
{\slowpokes}-II catalog. This distribution is strongly skewed towards
smaller separations, with 51\% of the pairs at 1$\arcsec$--7$\arcsec$ and 72\%
at 1$\arcsec$--10$\arcsec$.

\begin{figure}[hb]
  \begin{centering}
  \includegraphics[width=1\linewidth]{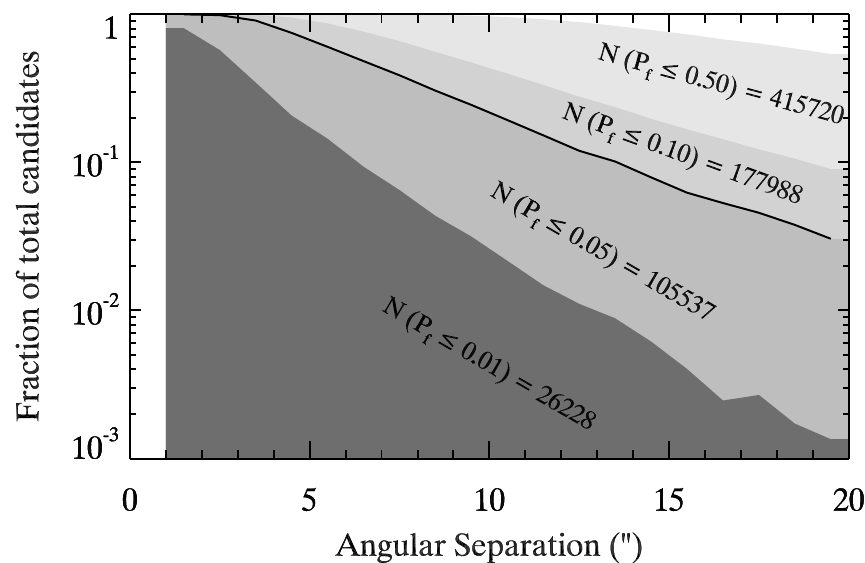}
  \caption{The fraction of binary candidates that meet the {\Pf}
    $\leq$ 1\%, 5\%, 10\%, and 50\% thresholds as a function of
    angular separation. The number of pairs that meet each threshold is
    also shown. We have chosen {\Pf} $\leq$ 5\% (dark line) for the
    {\slowpokes}-II catalog, resulting in 105,537 wide binaries.
  }
  \label{Fig:pf_threshold}
  \end{centering}
\end{figure}

Figure~\ref{Fig:pf_threshold} shows the distribution of the fraction
of binary candidates that pass various {\Pf} thresholds: $\leq$ 1\%, 5\%,
10\%, and 50\%, with the total number of candidates also
shown. Obviously, as we discussed in  \citetalias{Dhital2010}, the
choice of a threshold is subjective and depends on the need. For
example, if a followup study needed to minimize the number of false
positives, a sample with {\Pf} $\leq$ 1\% should be chosen. However, if
the study needed bright, rare binaries (e.g., WD$+$dM binaries), it
might need to tolerate the high number of false positives and select
candidates with {\Pf} $\leq$ 10\%. With such a large number of binary
candidates, there is a flexibility in the number and types of binary
systems that is available to followup studies.

\begin{figure}[tb]
  \begin{centering}
  \includegraphics[width=1\linewidth]{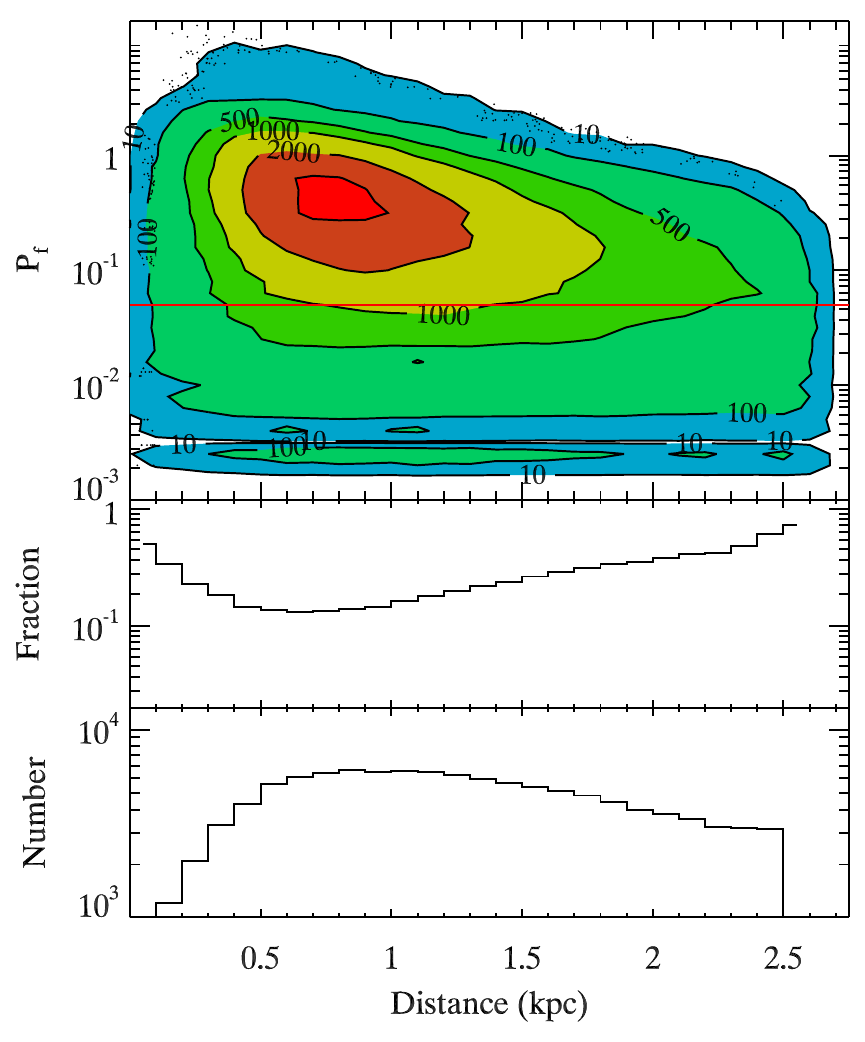}
  \caption{The number density, as represented by the count-ours, of the 514,424 binary candidates as a function of
    {\Pf} and distance. The red line is the threshold {\Pf}
    that we adopted for the {\slowpokes}-II catalog. The fraction of
    candidates that passed the {\Pf} threshold are shown in the middle
    panels while the distribution of the resultant 105,537 {\slowpokes}-II
    binaries are shown in the bottom panels.
  }
  \label{Fig:pf_dist}
  \end{centering}
\end{figure}

Figure~\ref {Fig:pf_dist} shows the number density of all candidate
pairs as a function of {\Pf} and photometric distance. In general,
{\Pf} is flat as a function of the distance, indicating that
the distance of a given pair has no significant effect on how likely
it is to be a chance alignment. More significantly, the number of
candidate pairs peaks around $\sim$800~pc and declines smoothly
afterwards. This is because an additional selection criteria, 
$\Delta~d \leq$ 100~pc (Eq.~\ref{Eq:pair_criteria}), which becomes
effective at large distances and rejects pairs with $\Delta~d$ within
1~$\sigma_{\Delta d}$ but larger than 100~pc. So even as the search
volume grows larger, the number of 
candidates actually decreases. However, the resulting candidates are
more likely to be binaries, as shown in the middle panel: the
fraction of candidate binaries that pass the {\Pf} $\leq$ 5\%
threshold reaches a minimum at $\sim$700~pc and increases sharply
afterwards. In fact, the acceptance rate for binaries at
$\sim$2200--2500~pc is higher than for binaries at any other
distance. At these distances, $\Delta~d \leq$ 100~pc implies that the
distances match within 6--7\% of each other, explaining the low rate
of false positives. Therefore, as shown in the lower panel of
Figure~\ref {Fig:pf_dist}, there are a relatively large number of
identified binaries at large distances.

\subsection{Distribution of binary separations}
\begin{figure}[htb]
  \begin{centering}
    \includegraphics[width=1\linewidth]{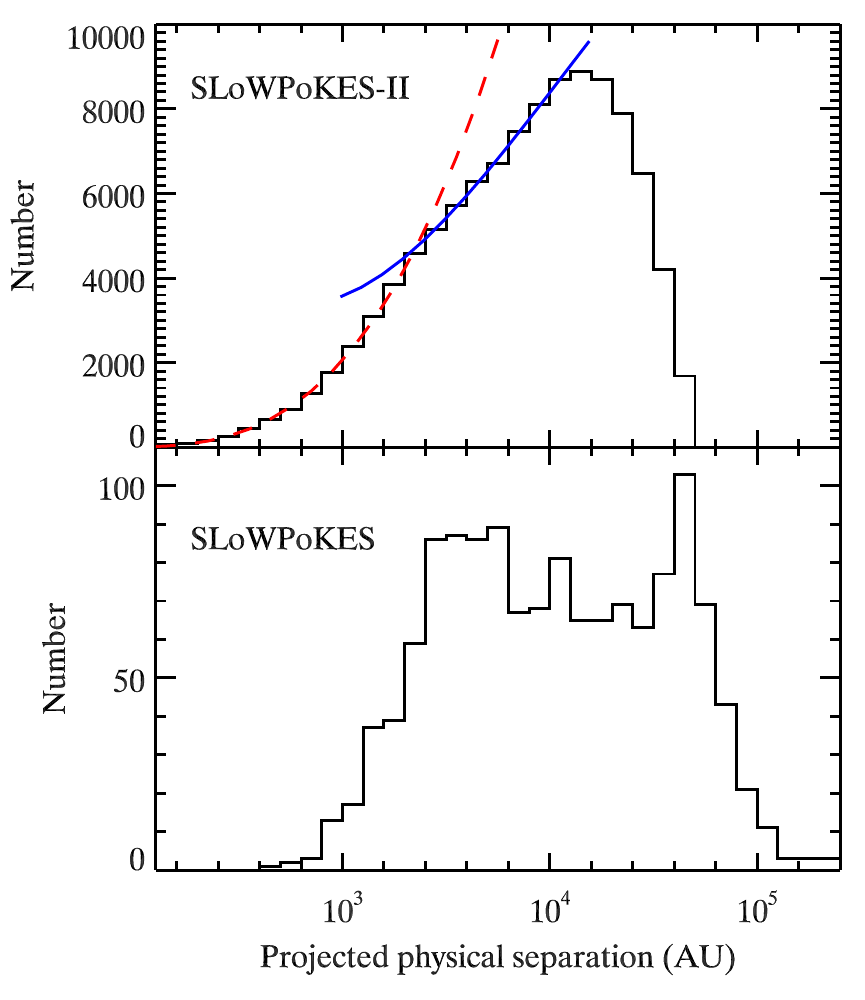}
    \caption{\protect\small
      The distribution of projected physical separations for the
      {\slowpokes}-II  binaries (top) and the {\slowpokes} CPM
      binaries (bottom). As {\slowpokes}-II was restricted to $\theta
      \lesssim 20\arcsec$, it only probes binaries smaller than
      $\sim$50,000~AU, thus missing the widest systems. Even so, the
      {slowpokes}-II distribution cannot be fit by a single functional
      form. We fit two polynomials with a break at $a = 10^{3.4}$~AU,
      shown in dashed red and solid blue lines in the figure. This
      inflection point suggests the presence of multiple populations
      of wide binaries.
    }
    \label{Fig:sep}
  \end{centering}
\end{figure}

Figure~\ref{Fig:sep} shows the distribution of projected physical
separations of the {\slowpokes} and {\slowpokes}-II binaries.\footnote{In
\citetalias{Dhital2010} we used a statistical correction to scale
the projected physical separation to semi-major axis; however,
\citet{Dupuy2011} have shown that the correction factor is
significantly smaller and highly dependent on the orbital
parameters. They calculated correction factors based for their sample
of $<$4~AU binaries. The sample presented here are in a completely
different separation regime, making it unlikely that the same
correction factors can be used. Therefore, we have eschewed semi-major
axis and chosen to work in projected physical separation space.}
The {\slowpokes} distribution exhibited a bimodal distribution, which
we interpreted as either the presence of two formation modes or the
preferential destruction of the widest pairs in the timescale of a
1--2~Gyr \citepalias{Dhital2010}. These binaries were 7--180$\arcsec$
systems, within an average distance of 1000~pc. In {\slowpokes}-II, we are not 
sensitive to binaries as wide, as the lack of proper motions limits us
to angular separations of 1--20$\arcsec$; however, the binaries are up
to 2500~pc away. Clearly, the selection biases and the resultant
distributions of the identified binaries involved are markedly
different. For example, the steep falloff of the {\slowpokes}-II
distribution at $a > 10^{4.1}$~AU is due to our insensitivity to
binaries at those separations.

In Paper~I we found a dip in the distribution of binary separations at
$a\sim 10^4$ AU (bottom panel of Figure \ref{Fig:sep}), but at first
glance this is not apparent in the new {\slowpokes}-II distribution (top
panel). However, there is an inflection point at $a \sim 10^{3.4}$~AU (2500~AU),
where the slope of the distribution becomes noticeably shallower. In
Figure~\ref{Fig:sep}, two power-law fits are shown on either side of
the inflection at $a \sim 10^{3.4}$~AU. This inflection is at a similar
separation scale as seen in previous samples of binary stars
\citep{Allen2000,Lepine2007a}, which were interpreted as wider binaries
being disrupted beyond a critical separation and being less common, as
was originally predicted by \citet{Opik1924}. However, it is much smaller
than the bimodality seen in \citetalias{Dhital2010} at $\sim$20,000~AU. 
Whether present at formation or sculpted later in life, the multiple
modes in binary populations is intriguing and suggests that interstellar
interactions are more common than thought.

\subsection{Binary mass distribution}
\begin{figure*}[htb]
  \begin{centering}
  \includegraphics[width=1\linewidth]{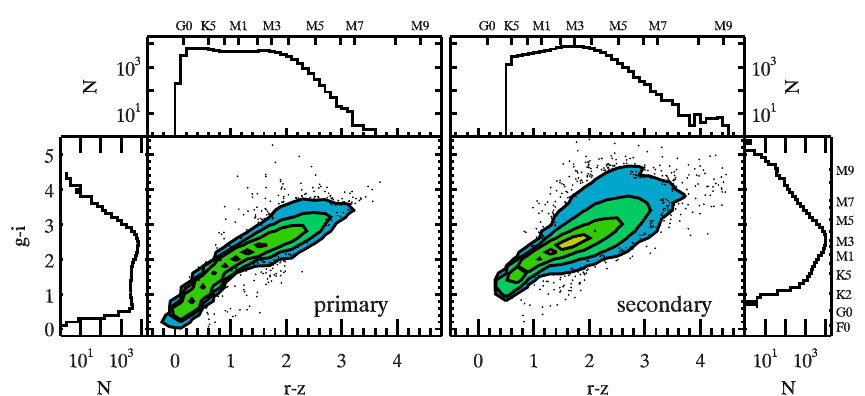}
  \caption{The $g-i$ and $r-z$ color distributions for the primary
    (left) and secondary (right) components of the {\slowpokes}-II
    binaries, with the contour levels at 10, 100, 1000, and 5000. The
    histograms of the distributions are plotted along 
    the sides, and the inferred spectral types are also shown. We
    defined the primary component to be have the bluer $r-z$ color.
    Both the primary and the secondary distributions show definitive
    peaks at $\sim$M2--M4 spectral types in both the $g-i$ and $r-z$
    colors. As our initial target sample used a cutoff on the $r-i$
    and $i-z$ colors at $\sim$K5 spectral type, there is a sharp
    cutoff at the blue end of the $r-z$ distribution for the secondary
    component.
  }
  \label{Fig:color}
  \end{centering}
\end{figure*}

\begin{figure}[hb]
  \begin{centering}
  \includegraphics[width=1\linewidth]{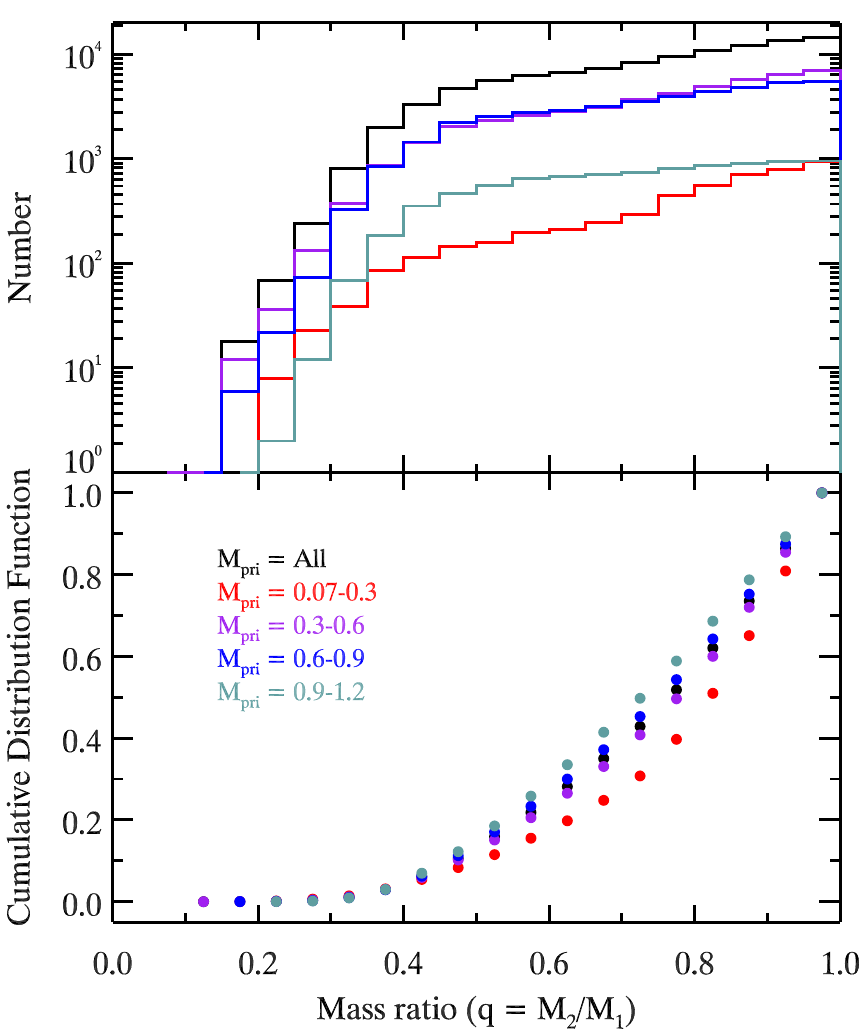}
  \caption{The mass ratio distribution of {\slowpokes}-II systems
    shown as histogram (top) and a cumulative distribution function
    (bottom). The black line and dots show the all the {\slowpokes}-II
    binaries while the red, purple, blue, and cyan show the
    distributions for systems with different primary masses.
    masses were calculated by interpolating from their $g-i$ and $r-z$
    colors using \citet{Kraus2007}; the minimum mass was set at
    0.075~{\Msun}. There is a large bias towards systems with similar
    masses, with the median mass ratio being $\sim$0.78.
    While the systems with larger primary masses seem more
    likely to have low mass ratios as compared to systems with
    low-mass primary, it is largely due to biases inherent in SDSS and
    our identification techniques.
  }
  \label{Fig:q}
  \end{centering}
\end{figure}

Figure~\ref{Fig:color} shows the $g-i$ and $r-z$ color
distributions for the primary (left) and secondary (right) components
of the {\slowpokes}-II binaries. The primary is defined as the
component with the bluer $r-z$ color. The histograms for the color
distributions are shown along the top and sides of each panel.
The inferred spectral types, based on the median color--spectral type
relations \citep{Covey2007,West2011}, are shown along the top axes. 
In general, the {\slowpokes}-II binaries reflect the SDSS low-mass
dwarf population in both $g-i$ and $r-z$ colors. The distributions
peak between the M2--M4 spectral types, same as the field mass
function \citep{Bochanski2010}. The primary color distributions also
exhibit a pileup at the bright end of the distribution, which is most
likely a selection bias. {\slowpokes}-II binaries span a large range
of colors (and masses). There are a significant number of primary
components from early--mid G to the mid-M dwarfs while the secondaries
extend from K5 to the M9 spectral types.  The blue end of the secondary component
distributions are defined by the color cuts imposed on our initial
target sample ($r-i \geq $0.3, $i-z \geq$ 0.2). While {\slowpokes}
contained only a handful of binaries after $\sim$M6--M7, there are now
a significant number of binaries at the low-mass end of the stellar
main sequence at large distances.

Figure~\ref{Fig:q} shows mass ratio distribution of {\slowpokes}-II
binaries as a histogram (top) and a cumulative distribution function
(bottom). The black line and dots represent the entire catalog while
the red, purple, blue, and cyan show distributions for different
primary masses. The masses were calculated by interpolating from their
$g-i$, $r-z$ colors, or $i-z$ based on \citet{Kraus2007}. We set the
lowest mass at 0.075~{\Msun}, the hydrogen burning minimum mass \citep{Burrows1997}, as the
$i-z$--mass relation is even less constrained at lower masses.
All of the distributions are skewed towards similar masses, with
notable deficit at the lowest mass ratios, below $q \sim$0.4. This is
mostly due to an additional selection criterion that at least one component be K5 or late
($\leq$0.70~{\Msun}), which means the lowest mass ratio possible for a
K5 primary is $\sim$0.11. Moreover, the steep mass--luminosity ratio among the
M spectral types means SDSS photometry would only detect a M6 or
earlier companion for the K5 primary, restricting the lowest possible
mass ratio to $\sim$0.17. Similarly, for a M4 primary
($\sim$0.20~{\Msun}), the observed mass ratio is always greater than 0.39. 
After $q \sim$0.4, the various cumulative distributions show a
relatively smooth progression that is similar for the different
masses. There is a preference for a larger $q$ at the lowest masses and
a smaller $q$ at the highest masses, but those are most likely due to
selection biases as discussed above.

\section{Discussion}\label{Sec:discussion}
We have assembled a large sample of 105,537 wide binaries with
projected physical separations of $\sim$1000--60,000~AU. The
identification was based on matching 3D position of stars in the SDSS
photometric sample and assessing the probability of chance alignment
with our Galactic model. All of the binaries in the {\slowpokes}-II
catalog have a probability of chance alignment of $\lesssim$5\%. For
comparison, the original {\slowpokes} contained 1342 CPM binaries with
separations of 7$\arcsec$--180$\arcsec$ (500--100,000~AU) at
$\sim$100--800~pc. While our initial aim was to combine these two samples
for a detailed analysis, the regimes they probed are much different. For example, despite
extending to $\gtrsim$2000~pc, the widest {\slowpokes}-II binary is a
factor of two smaller than the widest {\slowpokes} binary. Proper
motions allowed for the identification of CPM binaries up to
180$\arcsec$, whereas we were restricted to 20$\arcsec$ in this
paper. Thus, even though the two samples are are neither complete nor
continuous, they are complementary and are both of high fidelity.

\subsection{{\slowpokes-II}: A rich, diverse catalog of binaries}
In addition to being the largest catalog of wide binaries,
{\slowpokes}-II contains a diversity of systems---in mass, mass
ratios, metallicity, separations, evolutionary states, and Galactic
positions---that should facilitate followup studies to characterize
the properties of low mass stars. In addition, we have identified three
specific subsets of candidate binaries that are rare: (i) wide VLM
systems, (ii) WD$+$dM systems, and (iii) sdM$+$sdM binaries. 
They could prove uniquely useful in characterizing those
stellar types.

The diversity of systems should allow for in-depth probes of
different aspects of low-mass stellar physics. Specifically, the
flexibility offered by the large sample size and all-sky nature of
{\slowpokes}-II could be exploited by large spectroscopic surveys
where fibers are often available in certain parts of the sky. For
example, we were awarded 1000 fibers in the Baryon Oscillation 
Spectroscopic Survey of SDSS-III~\citep[BOSS;][]{Dawson2013} to
acquire spectra of 500 {\slowpokes}-II binaries. This was in addition
to the 1000 fibers awarded earlier for the {\slowpokes} binaries. We
are using the combined sample $\sim$1000 binaries to ascertain the
age--activity and metallicity relationships in low-mass M dwarfs
(Massey et al. \textit{in prep.}). 

\subsection{Multiple pathways for wide binary formation}\label{Sec:formation}
In \citetalias{Dhital2010} we noted the presence of a bimodality in
the physical separation distribution of the binaries (Figure~\ref{Fig:sep}).
When compared to the dynamical dissolution timescales
\citep{Weinberg1987}, the bimodality in the {\slowpokes} distribution
led us to suggest that it comprised of two different population of
binaries with a break at $\sim$20,000~AU: (1) a ``wide'' population that is dynamically stable over
$\sim$10~Gyr and (2) an ``ultra-wide'' population of young,
loosely-bound systems that will dissipate in 1--2~Gyr. In a followup
high-resolution imaging study with the Keck II and Palomar Laser 
Guide Systems with Adaptive Optics, we found that the frequency of a
close companion was higher in wide binaries than in single stars 
\citep{Law2010}. Moreover, the frequency increased significantly with
wide binary separation, suggesting that different formation
modes were at work.

In {\slowpokes}-II we have found further evidence of multiple
populations in the distribution of physical separations (Figure~\ref{Fig:sep}). 
While more subtle than in \citetalias{Dhital2010}, a break is
clearly evident at $\sim$2500~AU. This is much smaller than the
critical separation in \citetalias{Dhital2010} but approximately where
we saw the first peak in the {\slowpokes} sample. It is also at the same separation scale as
seen in previous studies \citep{Opik1924,Allen2000,Lepine2007a}.
That an inflection in the separation distribution stands out amongst
the myriad of selection biases and incompletenesses is quite
significant. Further investigation into these multiple populations and
their origins is clearly warranted. Whether wide binary distributions are
segregated at birth \citep[e.g.,][]{Reipurth2012} or sculpted as they
traverse around the Milky Way \citep[e.g.,][]{Weinberg1987,Jiang2010}
has important implications on our interpretation of observed binary
populations and, consequently, on our understanding of binary star formation. 

There are no signatures of multiple populations in the color (a proxy
for mass; Figure~\ref{Fig:color}) or mass ratio (Figure~\ref{Fig:q}) 
distributions. Even if there were such signatures were present,
identifying them from amongst the selection biases without further
data would be extremely difficult.
Recently, there has been extensive discussion on the formation
and stability of the extremely wide systems. This has partly been
motivated by the large samples of wide binaries that have become
available over the past decade
\citep[e.g.,][]{Chaname2004,Lepine2007a,Sesar2008,Dhital2010}. 
In addition, such wide binaries have been identified at very
young ages in Orion \citep*{Connelley2009} and in Taurus and Upper Sco
\citep{Kraus2009b}. Numerical simulations have explored mechanism that
would form wide binaries primordially or via dynamical
interactions, with three different pathways suggested. First,
observations of protostars at the end of long, extended filaments of
molecular gas \citep{Tobin2010} argues that wide binaries can
form with primordial separations of $\sim$0.1~pc. Second, primordial triple or
quadruple systems become extremely hierarchical by scattering or even
ejecting one of the components \citep{Reipurth2012}. Third,
simulations have shown wide binaries forming in the halo of clusters
via dynamical interactions or as they escape the cluster
\citep{Bate2005,Kouwenhoven2010,Moeckel2010,Moeckel2011}. These
pathways are unique and expected to imprint distinct signatures on the
resultant population. For example, the \citet{Reipurth2012} pathway
would produce a high frequency of triples and quadruples while
dynamical interactions would produce a very high proportion of
low-mass binaries.

There is no clear evidence from the {\slowpokes} samples that any of
these pathways are dominant. Rather, there is reason to believe all
of the pathways could be operational. Without a reliable age indicator
there is no way to observationally determine if components of an
wide binary formed together. In fact, it might be impossible to
do so even with an age indicator as the age difference of a few
million years can be indistinguishable once the stars are in the
Galactic field. We suggest that there is no principal pathway for wide
binary formation. Determining the relative efficiencies of the
different formation modes would be more valuable.

\subsection{Wide VLM binaries: Too numerous for the ejection hypothesis?}
{\slowpokes} did not contain wide very low-mass \citep[VLM; M
$\lesssim$ 0.1~{\Msun}][]{Burgasser2007c} or brown dwarf (BD) binaries.
They were too red and too faint to be detected in the USNO-B
survey and, therefore, did not have measured proper motions.
Finding more VLM/BD systems was one of our primary motivations for identifying binaries
without proper motions. Only eleven wide ($\gtrsim$100~AU) VLM/BD
systems are currently known, including a VLM triple at 820~AU
\citep{Burgasser2012} and binaries at 5100~AU \citep{Artigau2007} and
6700~AU \citep{Radigan2009}. With the formation processes and
techniques to measure properties for VLM/BDs not completely
understood, the value of a larger sample could not be overstated.

Until recently, the processes by which very low mass \citep[VLM; M
$\lesssim$ 0.1~{\Msun}][]{Burgasser2007c} stars and brown dwarfs (BDs)
form were thought to be different from those of higher-mass
stars. \citet{Reipurth2001} suggested that VLM/BDs are protostars
that are ejected from their natal cores before they can accrete and grow to
stellar masses. Such ejections were seen in numerical simulations
\citep[e.g.,][]{Bate2002a}, and all observed VLM/BD binaries had
enough binding energy to have survived such ejections
\citep{Burgasser2003a,Close2003}. However, with the discovery of each
subsequent VLM/BD binary wider than $\sim$100~AU, the viability of the
ejection hypothesis had been called into question
\citep[e.g.,][]{Dhital2011,Burgasser2012}.

\begin{figure}[htb]
  \begin{centering}
  \includegraphics[width=1\linewidth]{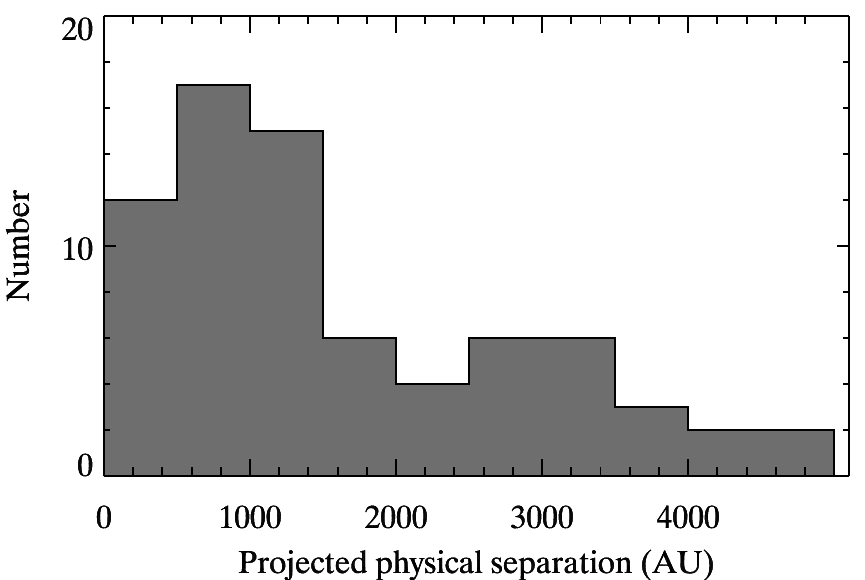}
  \caption{
    Distribution of {\slowpokes}-II wide binaries, at the end of the
    main sequence. These were selected by requiring $i-z \geq 1.14$,
    the median color for a M8 dwarf \citep{West2011}. As only eleven
    wide VLM binaries are currently known, a larger population has
    strong implications on how we interpret star formation at the
    lowest masses.
  }
  \label{Fig:vlm}
  \end{centering}
\end{figure}

Figure~\ref{Fig:vlm} shows the projected physical separation
distribution for a sample of 141 binaries with $i-z \geq 1.14$, the 
median color of an M8 dwarf \citep{West2011}.  While spectroscopic
followup are ongoing to confirm their spectral types (a proxy for
mass), it seems clear that significant number of systems with 
separations of 1000--5000~AU exist in the Galactic field. We note that
photometric colors can be particularly deceiving in the VLM/BD regime,
so the need for spectra cannot be overemphasized. However, such a
large number of in wide VLM binary candidates can inform us a lot about
their formation. \footnote{For reference, the current sample of all VLM binaries
is $\gtrsim$120 (T. Dupuy, \textit{priv. comm.}).} These
systems most definitely did not form primordially via ejection, as their
binding energy is too small to have survived a dynamical
kick. With 11 previous systems and 141 candidates presented here, we
cannot explain away the wide VLM binary population as exceptions.

The alternative pathway to VLM/BD formation is that they form in a
manner similar to stars via gravoturbulent fragmentation
\citep{Hennebelle2011,Jumper2013}. The detection of the first pre-BD core,
\textit{Oph~B-11} \citep{Andre2012}, has helped further that idea
significantly. Our sample does not allow for rigorous testing of this
hypothesis; we will wait for followup observation that will better
characterize the binaries. However, neither the separation distribution nor the mass
ratio distribution show any significant change as a function
of mass. While the systems with less massive primaries are more likely
to be equal-mass (Figure~\ref{Fig:q}). This characteristic is a bias of our sample mostly
due to insensitivity to lower-mass, fainter companions and to setting
the minimum mass at 0.075~{\Msun}. The trend of more equal-mass
binaries is also a gradual change at all masses, with no significant
break at the VLM/BD regime. 

There is no suggestion of multiple populations of wide VLM binaries,
as was observed for wide stellar binaries in Section~\ref{Sec:formation}. 
However, we cannot discount the possibility that these wide
VLM binaries were bound post-formation
\citep{Kouwenhoven2010,Moeckel2010,Moeckel2011}. Numerical
simulations show that a molecular core typically fragments into
3--5 cores, whence the smaller cores are ejected before they can grow
into stellar masses \citep{Bate2012}. Open clusters could have
hundreds of these VLM protostars floating around, increasing the
likelihood of multiple ones interacting at the same time and forming
wide, gravitationally-bound systems. The efficiency of these
capture-like processes could be much higher in the VLM/BD regime.

\section{Conclusion}\label{Sec:conclusions}
We have identified the {\slowpokes}-II catalog of 105,537 wide,
low-mass binaries without using proper motions. While false positives
are inherent in all statistical samples, we have required a relatively
stringent probability of chance alignment, {\Pf}, as calculated by our
Galactic model to be $\leq$5\% for each of our binaries. Most binaries
have a much smaller probability of chance alignment. The entire sample
is expected to have 2464 (2.3\%) chance alignments. In addition, we
have identified 944 sdM$+$sdM and 450 WD$+$dM candidate binaries, which also
have a $\leq$5\% probability of chance alignment. However, as their
identification as an sdM or WD is based on photometry alone, there is
a higher probability of them being false positives.

The \textsc{STAR} table in SDSS suffers from a $\sim$6\%
incompleteness. In particular, $\sim$6\% of the spectroscopically
confirmed, \textit{bona fide} M dwarfs \citep{West2011} are missing
are not included in \textsc{STAR}. Instead, they are classified as
extended sources due to the presence of a nearby or partially
resolved stellar object. This incompleteness could be exaggerated for
the spectroscopic sample as a significant number of M dwarfs were
observed as potential galaxies and quasars \citep{Adelman-McCarthy2006}. 
We advise caution when using the \textsc{STAR} table, especially for
studies that seek to do a complete census or build a complete sample
of stars in the Galaxy.

{\slowpokes}-II binaries have projected physical separations of
500--50,000~AU at distances up to 2500~pc. We are not sensitive to
binaries as wide as the ones identified in \citepalias{Dhital2010} as
only searched within 20$\arcsec$. Given the dynamic magnitude limits
of the SDSS, our sample is likely largely dominated by a confluence of
selection biases making is hard to construct or even complete
samples. For example, the color distributions for both the primary and
secondary components show a higher proportion of G- and K-type dwarfs, as
compared to the field population \citep{Bochanski2010}. This is likely
because they are brighter and more likely to have been detected. With
complete samples not feasible, the main contribution of
{\slowpokes}-II is the large, diverse sample of binaries that will
facilitate detailed studies of low-mass star properties. Sub-samples
are already part of various studies to probe the age--activity,
rotation--age, and metallicity of M dwarfs. Most importantly,
{\slowpokes}-II contains a significant number of binaries at the
late-M spectral types, enabling the studies to probe into the very
bottom of the main sequence.

The distribution of binary physical separations exhibits a marked
inflection which can be interpreted as representing two binary
populations, consistent with our results in \citetalias{Dhital2010}
and \citet{Law2010}. While the incompleteness in
our sample and lack of any way to quantify or correct for that
incompleteness prevents us from characterizing the populations, the
separation distribution exhibits a marked inflection which can be fit
by two more polynomials. Combined with our results in
\citetalias{Dhital2010} and \citet{Law2010}, we infer that the wide
binary population in the Galactic field is composed of multiple
different populations, either from different formation mode or via
different dynamical histories.

We have identified 141 wide binaries in which both components are
VLMs. This is 7$\times$ larger than the current sample of wide, VLM
binaries. While spectroscopic data are needed to confirm their VLM
status, it is becoming clear that wide VLM binaries are not
exceptions. These wide systems are critical in understanding VLM/BD
formation, as their binding energies are too low to have formed via
the ejection mechanism \citep{Reipurth2001}. As searches go deeper
and identify more wide, VLM binaries, it looks unlikely that VLM/BDs
are formed primarily via ejection. The data indicate that there is no
change in formation mechanism with stellar mass, suggesting VLM/BDs are
also formed via gravoturbulent fragmentation, like their more massive counterparts
 \citep{Hennebelle2011,Jumper2013}.

The {\slowpokes} and {\slowpokes}-II catalogs are available on
the Filtergraph portal$^5$.
\citep{Burger2013}. The portal allows for dynamic plotting of these
large data sets and real-time filtering of the data using
user-specified criteria.  It is useful for target selection as its smoothly
transitions between graphical and tabular visualizations of the data
and allows for selection of data points with the keyboard or the mouse.

Lastly, we have demonstrated a methodology to identify bona fide wide
binaries using just photometry together with a galactic model to
assess false positives. Really, this technique harks back to
\citet{Michell1767}'s discovery of ``double'' stars, where he argued
that some pairs of stars were too close to each other, as compared to
mean distances between stars, to be unrelated. We have used the
all-sky data from SDSS and our understanding of the Galaxy's stellar
distribution to statistically identify bona fide binary stars without
the advantage of proper motions.

\acknowledgements
The authors would like to thank Jeff Andrews, Matthew Bate, Adam
Burgasser, Ben Burningham, Rebecca Oppenheimer, and Leigh Smith for useful discussions and
Dan Burger for help with setting up the online data visualization portal.

SD, AAW, and KGS acknowledge funding support through NSF grant
AST--0909463. AAW also acknowledges support through AST--1109273. 

This work was conducted in part using the resources of the Advanced
Computing Center for Research and Education at Vanderbilt University,
Nashville, TN. 

Funding for the SDSS and SDSS-II was provided by the Alfred P. Sloan
Foundation, the Participating Institutions, the National Science
Foundation, the U.S. Department of Energy, the National Aeronautics
and Space Administration, the Japanese Monbukagakusho, the Max Planck
Society, and the Higher Education Funding Council for England. The
SDSS was managed by the Astrophysical Research Consortium for the
Participating Institutions.

We acknowledge use of the ADS bibliographic service.

\bibliography{ads}

\end{document}